\documentclass[aps,pra,twocolumn,amsmath,amssymb]{revtex4-2}
\usepackage{graphicx}
\usepackage{bm}% bold math
\usepackage{multirow}
\usepackage{ulem}
\usepackage{color}

\usepackage{comment}
\usepackage{braket}
\usepackage{adjustbox}
\usepackage{floatrow}

\usepackage{amsmath}

\usepackage{longtable}

\usepackage{hyperref}

\begin{document}

\title{{B-Spline for Self‑Consistent Field Theory with a $Z$‑Dependent Pauli Potential for  Atomic Binding Energies}}

\author{$^{1}$Vipul Badhan}
\author{$^2$Russell B. Thompson}
\author{$^{1}$Bindiya Arora}
\email{bindiya.phy@gndu.ac.in}
\affiliation{$^1$Department of Physics, Guru Nanak Dev University, Amritsar, Punjab 143005, India}
\affiliation{$^2$ Department of Physics \& Astronomy and Waterloo Institute for Nanotechnology, University of Waterloo, 200 University Avenue West, Waterloo, Ontario, Canada N2L 3G1}

\begin{abstract}
Polymer self-consistent field theory (SCFT) has recently been established as a promising alternative framework to Kohn–Sham density functional theory (KS-DFT) for modeling quantum many-body systems. 
It uses real-valued propagators instead of orbitals, simplifying the self-consistent numerical solution. 
However, SCFT suffers from inaccuracies in heavy-element systems due to the approximate treatment of the Pauli potential, particularly the use of a constant repulsion strength parameter. In this work, we address this central limitation by introducing a $Z$-dependent Pauli potential that improves agreement with Hartree-Fock (HF) results. Furthermore, we advance SCFT implementation by employing B-spline basis functions-highly localized, piecewise-polynomial functions widely used in atomic structure theory. We demonstrate that B-splines provide a flexible and efficient representation of electronic structure, and present results for atomic binding energies from hydrogen to xenon. Comparisons with HF theory and prior SCFT calculations using Gaussian basis sets highlight the improved accuracy achieved with the $Z$-dependent potential. 
\end{abstract}

\maketitle

\section{Introduction}~\label{intro}

The accurate and efficient simulation of quantum many-body systems remains a central challenge in computational physics and chemistry. While first-principles methods like quantum Monte Carlo offer high accuracy, their computational cost often limits their application to small systems. For larger systems, density functional theory (DFT) \cite{hohenberg1964, kohn1965} has emerged as the dominant framework due to its favorable balance between accuracy and computational efficiency. In particular, Kohn-Sham DFT (KS-DFT) achieves chemical accuracy in many real-world applications by mapping the interacting electron system onto a non-interacting system with the same density, solved through single-particle orbital functions.

In recent years, polymer self-consistent field theory (SCFT) has been effectively applied to quantum many-body systems as an alternative computational framework \cite{Thompson2020, Sillaste2022, LeMaitre2023a, LeMaitre2023b}. Originating in polymer physics, SCFT employs a statistical field theory approach that represents quantum particles as continuous paths. {The method offers potential {numerical} advantages, as its equations are parabolic initial-value problems that are numerically simpler to solve than the elliptic boundary-value problems characteristic of KS-DFT.} Rather than using orbital functions, SCFT employs real-valued propagators as solutions to modified diffusion equations that represent quantum statistical weights.

Studies have established that the equations governing SCFT are mathematically equivalent to those of KS-DFT \cite{Thompson2019}, ensuring, through the core principles of DFT, that SCFT aligns with the predictions of quantum mechanics. {Beyond their mathematical equivalence, SCFT offers several formal advantages. By employing propagators governed by initial-value equations rather than solving Kohn–Sham orbital eigenvalue problems, SCFT avoids explicit orbital representations, and its structure may enable greater parallelization.} Additionally, deriving SCFT equations from the classical partition function framework provides novel perspectives on foundational quantum mechanical concepts \cite{Thompson2022, Thompson2023, Kealey2024, Thompson2025}.

However, the simplified theoretical framework of SCFT introduces approximations that can lead to inconsistencies, particularly for heavy elements. As noted in Ref.~\cite{LeMaitre2023a}, these limitations stem primarily from the approximate treatment of the Pauli exclusion principle through a Pauli potential with a constant strength phenomenological parameter $g_0^{-1}$. Recent work~\cite{Thompson2025} suggests that substantial improvements could be achieved by refining this Pauli potential, specifically by making its strength dependent on the electron number $Z$ rather than treating it as a fixed parameter \cite{LeMaitre2023a}. In the present work, we address this challenge by developing and testing a $Z$-dependent Pauli potential formulation.

The choice of basis functions represents another critical aspect of SCFT implementation. Like KS-DFT, the computational expense of SCFT scales cubically with the number of basis functions. Previous implementations have employed orthogonal basis sets for simplicity \cite{Thompson2019, Thompson2020, Sillaste2022} and more recently, optimized non-orthogonal Gaussian basis sets tailored for atomic calculations \cite{LeMaitre2023a, LeMaitre2023b}. An important direction for advancing the computational framework of SCFT involves exploring more efficient basis sets that balance accuracy with computational cost.

In this context, B-splines present a particularly promising alternative. Widely adopted in atomic physics for their numerical stability and efficiency, B-splines offer a versatile framework for representing wavefunctions through piecewise polynomial functions characterized by strong localization and recursive definitions \cite{johnsonsplines,Sapirstein_1996}. Their success in traditional electronic structure calculations motivates their application to SCFT to improve practical aspects of the calculation. 

  {The numerical implementation of the SCFT equations in the present work focuses on single-atom systems, where spherical Bessel functions have been successfully employed~\cite{Thompson2022}. For such systems, a B-spline basis provides an accurate and efficient representation of the propagator and fields while respecting spherical symmetry. Although diatomic molecules have been treated using Fourier and cylindrical Bessel bases~\cite{Sillaste2022}, and B-splines have been proposed as a flexible alternative for polyatomic systems with lower symmetry~\cite{TOFFOLI200225,TOFFOLI2024,Toffoli20242}, the extension to polyatomic geometries is left for future work. Nevertheless, the favorable properties of B-splines such as localized resolution, systematic convergence, and efficient treatment of boundary conditions, make them well-suited for ultimately solving the modified diffusion equation in molecular geometries. Within the scope of this paper, however, we restrict ourselves to single-atom calculations using a B-spline basis.}

This paper presents a dual advancement in SCFT methodology: the implementation of B-spline basis functions and the development of a $Z$-dependent Pauli potential. We assess these improvements through systematic calculations on atomic systems ranging from hydrogen to xenon. The remainder of this paper is organized as follows. Section 2 provides a concise review of the SCFT framework and details our B-spline implementation. In Section 3, we present binding energy calculations that benchmark our approach against Hartree-Fock (HF) theory \cite{hf} and previous Gaussian-basis SCFT results with constant Pauli potentials \cite{LeMaitre2023b}. We demonstrate that our $Z$-dependent Pauli potential yields significant improvements in accuracy. 
Finally, Section 4 summarizes our conclusions and discusses potential directions for future research.

\section{Theory}~\label{theory}

\subsection{Self-Consistent Field Theory for Quantum Systems}

The application of polymer self-consistent field theory to quantum many-body systems originates from the quantum-classical isomorphism \cite{Chandler1981,Feynman1953b}, where quantum particles can be represented as ring polymers in a four-dimensional thermal space with imaginary time dimension $\beta = 1/(k_B T)$. According to this picture, each electron follows a thermal world-line or trajectory along the $\beta$ axis, forming a closed contour (ring polymer) as required by Feynman's path integral formulation \cite{Feynman}. In this representation, the quantum statistical mechanics of electrons maps onto the classical statistical mechanics of Gaussian threads \cite{Thompson2022}.

For a system of electrons in an external potential, the canonical ensemble free energy within the SCFT framework is given by \cite{Thompson2020,LeMaitre2023a}:

\begin{eqnarray}
F[\{n\},\{w\}] &=& -\frac{1}{\beta}\sum_{i}N_{i}\ln Q_{i}(\beta) \nonumber \\
               &-& \sum_{i}\int d\mathbf{r}\, w_{i}(\mathbf{r},\beta)n_{i}(\mathbf{r},\beta) + U[\{n\}],
\end{eqnarray}
where $n_{i}(\mathbf{r},\beta)$ is the electron density of the $i$th electron pair at position $\mathbf{r}$, imaginary time $\beta$, $N_i$ is the number of electrons in the $i$th pair, $w_{i}(\mathbf{r},\beta)$ are conjugate fields associated with each density pair, $Q_{i}(\beta)$ are single-particle partition functions for each electron pair, and $U[\{n\}]$ is the total potential. The argument $\{n\}$ indicates functional dependence on the set of all electron pair densities.   {This formulation assumes a restricted, spin-unpolarized description of the system, wherein electrons are treated in pairs.}

The total potential comprises of three contributions:

\begin{equation}
U[\{n\}] = U_{\rm ext}[n] + U_{\rm ee}[\{n\}] + U_{\rm P}[\{n\}],
\end{equation}

where $U_{\rm ext}[n]$ is the nuclear attraction, $U_{\rm ee}[\{n\}]$ is the electron-electron Coulomb repulsion, and $U_{\rm P}[\{n\}]$ is the Pauli exclusion potential. Note that in this formulation, exchange and correlation effects are not included however, these effects are intended to be captured through the Pauli potential and the statistical treatment of electron pairs.

\subsubsection{Nuclear Potential}

Treating the nucleus as a point particle ($\rho_{\rm ion}(\mathbf{r}) = \delta(\mathbf{r})$), the nuclear potential is given by:

\begin{equation}
U_{\rm ext}[n] = -Z\int\int n(\mathbf{r},\beta)V(|\mathbf{r}-\mathbf{r}'|)\rho_{\rm ion}(\mathbf{r}')d\mathbf{r}d\mathbf{r}',
\end{equation}

where $n(\mathbf{r},\beta) = \sum_i n_i(\mathbf{r},\beta)$ is the total electron density, $Z$ is the atomic number for neutral atoms, and $V(|\mathbf{r}-\mathbf{r}'|) = 1/|\mathbf{r}-\mathbf{r}'|$ is the Coulomb potential.

\subsubsection{Electron-Electron Interaction}

The electron-electron interaction is modeled using a Fermi-Amaldi-type expression that exactly accounts for self-interactions when dealing with electron pairs \cite{Ayers2005}:

\begin{eqnarray}
U_{\rm ee}[\{n\}] &=& \frac{1}{2}\sum_{ij}\left(1-\frac{\delta_{ij}}{N_i}\right)\nonumber \\
&&\int\int n_i(\mathbf{r},\beta)V(|\mathbf{r}-\mathbf{r}'|)n_j(\mathbf{r}',\beta)d\mathbf{r}d\mathbf{r}',
\label{ee}
\end{eqnarray}

where $\delta_{ij}$ is the Kronecker delta. This formulation exactly cancels self-interactions only for electron groups with one or two electrons. Using the shell approximation (more than two electrons per group) makes eq.\ref{ee} approximate, reducing it to a Fermi-Amaldi-type term as an approximation.

\subsubsection{Pauli Exclusion Potential}

To account for the fermionic nature of electrons, a Pauli exclusion potential is introduced. In the thermal-world-line picture, this corresponds to a repulsive pseudo-potential acting between infinitesimal segments of different ring polymer contours. Following Thompson \cite{Thompson2020}, and accounting for spin by grouping electrons in pairs, the Pauli potential is approximated as:

\begin{equation}
U_{\rm P}[\{n\}] = \frac{1}{2}\sum_{ij}(1-\delta_{ij})g_0^{-1}\int n_i(\mathbf{r},\beta)n_j(\mathbf{r},\beta)d\mathbf{r},
\end{equation}

where $g_0^{-1}$ is a parameter controlling the strength of Pauli repulsion, analogous to the excluded-volume parameter in polymer physics \cite{Edwards1965}. The factor of $1/2$ corrects for double counting. This expression represents an approximation where repulsion is averaged over all imaginary time slices rather than acting only between equivalent slices, which tends to overestimate the true Pauli repulsion \cite{LeMaitre2023a}.

\subsection{Diffusion Equations and Self-Consistency}

Variation of the free energy functional with respect to the fields and densities yields the self-consistent field equations. The central equations are modified diffusion equations for the propagators $q_i(\mathbf{r}_0, \mathbf{r}, s)$:

\begin{equation}
\frac{\partial q_i(\mathbf{r}_0, \mathbf{r}, s)}{\partial s} = \frac{\hbar^2}{2m}\nabla^2 q_i(\mathbf{r}_0, \mathbf{r}, s) - w_i(\mathbf{r},\beta)q_i(\mathbf{r}_0, \mathbf{r}, s),\label{prop}
\end{equation}

subject to the initial condition:

\begin{equation}
q_i(\mathbf{r}_0, \mathbf{r}, 0) = \delta(\mathbf{r} - \mathbf{r}_0).
\end{equation}

The total field $w_i(\mathbf{r})$ acting on the $i$th pair is composed of three contributions:

\begin{equation}
w_i(\mathbf{r}) = w_i^{\rm P}(\mathbf{r},\beta) + w_i^{\rm ee}(\mathbf{r},\beta) + w^{\rm ext}(\mathbf{r}),
\end{equation}

where

\begin{align}
w_i^{\rm P}(\mathbf{r},\beta) &= g_0^{-1}\sum_{j \neq i} n_j(\mathbf{r},\beta), \\
w_i^{\rm ee}(\mathbf{r},\beta) &= \int\left[n(\mathbf{r}',\beta) - \frac{n_i(\mathbf{r}',\beta)}{N_i}\right]V(|\mathbf{r}-\mathbf{r}'|)d\mathbf{r}', \\
w^{\rm ext}(r) &= -Z\int V(|\mathbf{r}-\mathbf{r}'|)\rho_{\rm ion}(\mathbf{r}')d\mathbf{r}'.
\end{align}

The single-particle partition functions and densities are then given by:

\begin{align}
Q_i(\beta) &= \int q_i(\mathbf{r}, \mathbf{r}, \beta) d\mathbf{r}, \\
n_i(\mathbf{r},\beta) &= \frac{N_i}{Q_i} q_i(\mathbf{r}, \mathbf{r}, \beta).
\label{ni}
\end{align}

These equations are mathematically equivalent to those of Kohn-Sham density functional theory \cite{Thompson2019}, ensuring that SCFT predictions are consistent with quantum mechanics through the theorems of DFT \cite{hohenberg1964, kohn1965,mermin}.

\subsection{Spectral Expansion with Non-Orthogonal Basis Sets}

The double spatial dependence of the propagators in equation~\ref{prop} makes real-space methods impractical. Instead, we employ a bilinear spectral expansion using basis functions $\{f_i(\mathbf{r})\}$:

\begin{align}
g(\mathbf{r}) &= \sum_i g_i f_i(\mathbf{r}), \\
g(\mathbf{r}_0, \mathbf{r}) &= \sum_{i,j} g_{ij} f_i(\mathbf{r}_0) f_j(\mathbf{r}),
\end{align}

where $g(\mathbf{r})$ and $g(\mathbf{r}_0, \mathbf{r})$ are arbitrary one- and two-variable spatial functions, respectively, and $g_i$, $g_{ij}$ are expansion coefficients.

Expanding the SCFT equations in terms of a non-orthogonal basis set introduces three fundamental molecular integrals:

\begin{align}
S_{ij} &= \int f_i(\mathbf{r}) f_j(\mathbf{r}) d\mathbf{r}, \\
L_{ij} &= \int f_i(\mathbf{r}) \nabla^2 f_j(\mathbf{r}) d\mathbf{r}, \\
\Gamma_{ijk} &= \int f_i(\mathbf{r}) f_j(\mathbf{r}) f_k(\mathbf{r}) d\mathbf{r},
\end{align}
which represent the overlap matrix, Laplacian matrix, and Gamma tensor, respectively. For spherical systems, we consider only radial dependence, following the finding that angular dependence has minimal effect on neutral atom electron densities \cite{Chowdhury2021}.   {We computed the necessary integrals using a modified trapezoidal rule with endpoint corrections, described using quadrature formula given in ~\cite{johnsonbook}. This type of quadrature formula maintains the simplicity of the trapezoidal rule, but can be made systematically more accurate. We chose 7 endpoint corrections in our calculations.}

The spectral representation transforms the diffusion equation into a matrix equation:

\begin{equation}
\frac{d\mathbf{q}_i}{ds} = \mathbf{S}^{-1}\mathbf{A}_i\mathbf{q}_i,
\end{equation}

where $\mathbf{A}_i = \frac{1}{2}\mathbf{L} - \boldsymbol{\Gamma}\mathbf{w}_i$, with $\mathbf{w}_i$ representing the field components for the $i$th pair. {The matrices and vectors are written in a boldface notation to suppress indices.} The solution is obtained through a generalized eigenvalue problem involving the matrix pencil $(\mathbf{A}_i, \mathbf{S})$.

\subsection{B-Spline Basis Functions}
While previous work has employed Gaussian basis functions for SCFT calculations \cite{LeMaitre2023a}, in this study we implement B-spline basis functions. B-splines are  piecewise polynomial functions constructed recursively from a set of knot vectors \( t_i \) distributed non-uniformly. The first and last $k$ boundary knots are defined as \cite{johnsonsplines,vbbasis}
\begin{eqnarray}
t_1 = t_2 = \cdots = t_k &=& 0, \nonumber\\
t_{n_{\rm max}} = t_{n_{\rm max}+1} = \cdots = t_{n_{\rm max}+k} &=& r_{\rm max},
\end{eqnarray}
where \( n_{\rm max} \) denotes the total number of splines and \( k \) is the order of spline. The interior knots are placed exponentially according to
\[
t_i = t_{k+1} e^{(i-k-1)h}, \qquad i = k+1,\dots,n_{\rm max}-1.
\]
The first nonzero knot \( t_{k+1} \) is set to \( 10^{-4} \), and the spacing parameter \( h \) is given by
\begin{equation}
h = \frac{\log(r_{\rm max}/t_{k+1})}{k + n_{\rm max} - 1}.
\end{equation}

The radial basis functions are then taken as
\begin{equation}
f_i(r) = B_{i,k}(r), \label{bbasis}
\end{equation}
where \( B_{i,k}(r) \) are defined via the Cox-de Boor recurrence formula
\begin{equation}
B_{i,1}(r) = 
\begin{cases} 
1, & t_i \le r < t_{i+1}, \\
0, & \text{otherwise},
\end{cases}
\end{equation}
and for \( k > 1 \)
\begin{eqnarray}
B_{i,k}(r) &=& \frac{r-t_i}{t_{i+k-1}-t_i}B_{i,k-1}(r) 
           + \frac{t_{i+k}-r}{t_{i+k}-t_{i+1}}B_{i+1,k-1}(r),\nonumber\\
&\qquad& i = 1,\dots,n_{\rm max}.
\end{eqnarray}

To make sure of the correct asymptotic behavior of the electron densities at the boundaries, the first and last splines are omitted from the basis expansion. The exponential distribution of knot yields a higher density of basis functions near the nucleus, where the electron density varies rapidly in the asymptotic region. The radial domain extends to \( r_{\rm max} = 110\) atomic units, which is sufficient to capture the asymptotic decay of the electron densities for all elements considered.

\subsection{Z-Dependent Pauli Potential}

A key limitation of the approximate Pauli potential in equation (5) is the use of a constant $g_0^{-1}$ parameter. As noted in Ref.~\cite{LeMaitre2023a}, this leads to inconsistencies, particularly for heavier elements where the approximation overestimates Pauli repulsion. To address this, we propose a $Z$-dependent Pauli potential strength:

\begin{equation}
g_0^{-1}(Z) = \frac{192}{Z},
\end{equation}

where $Z$ is the atomic number. This functional form is motivated by the need to reduce Pauli repulsion strength for heavier elements while maintaining accurate results for light elements. The specific numerical factor 192 is chosen based on systematic testing to optimize agreement with Hartree-Fock reference data across the periodic table.

This modification represents an improvement over the constant $g_0^{-1} = 10$ used in previous work \cite{LeMaitre2023a}, while maintaining the simple analytical form suitable for efficient computation. The $Z$-dependence captures the essential physics that Pauli repulsion effects scale differently with electron number than the simple pairwise approximation suggests.

\subsection{Computational Implementation}

The self-consistent solution proceeds as follows:
\begin{enumerate}
    \item Initialize electron density components using a Thomas-Fermi or other suitable approximation.
    \item Compute the molecular integrals $S_{ij}$, $L_{ij}$, and $\Gamma_{ijk}$ for the chosen B-spline basis set.
    \item For each electron pair $i$, construct the matrix $\mathbf{A}_i$ and solve the generalized eigenvalue problem.
    \item Compute the propagator matrices $\mathbf{q}_i$ and partition functions $Q_i$.
    \item Update the density components $\mathbf{n}_i$.% using equation \ref{ni}.
    \item Calculate the fields $\mathbf{w}_i$ and check for convergence using a density-weighted norm.
    \item Iterate until convergence is achieved.
\end{enumerate}

All calculations are performed with $\beta = 100$. The basis set convergence is monitored by varying the number of B-splines and verifying that results stabilize within acceptable tolerances.

\section{Results and Discussion}~\label{result}

Table~\ref{tab:binding_energies} presents atomic binding energies for elements from hydrogen through xenon, calculated using two different choices of the \( g_0^{-1} \) parameter in the SCFT formalism: a constant value \( g_0^{-1} = 10 \), and a \( Z \)-dependent value \( g_0^{-1} = \frac{192}{Z} \).  {The notation for atomic configurations in Table~1 follows the spectroscopic form given in Ref.~\cite{Thompson2020}. For each element, the configuration is built from the noble gas core of the previous period, followed by the valence subshells with numbers inside parenthesis indicating the number of electrons occupying that subshell. For example, $1s{(1)}$ denotes a single electron in the $1s$ orbital (hydrogen), $1s{(2)}$ denotes a filled $1s$ subshell (helium), and $[\text{He}]2s{(1)}$ denotes a helium core plus one electron in the $2s$ orbital (lithium).  Due to the spherical averaging approximation employed for atoms heavier than beryllium~\cite{Thompson2020}, electrons in nonspherical subshells (e.g., $2p$, $3p$) are lumped together with their spherical shells to reduce computational complexity. For example, instead of solving three separate diffusion equations for boron ($1s(2) 2s(2) 2p(1)$), we solve only two: one for the $1s(2)$ pair and another for the three $2s(2) 2p(1)$ electrons. This approximation is expected to be reasonable, and the errors introduced are small, as shown in Table~\ref{tab:binding_energies}} The Hartree-Fock reference values~\cite{hf}, along with the percentage deviations of the SCFT results from these HF benchmarks, are also provided. For context, values from Ref.~\cite{LeMaitre2023a}, obtained using SCFT with Gaussian basis functions, are included in brackets for elements H to Kr. All energies are reported in atomic units.

The core objective of this study is to systematically evaluate the accuracy of SCFT and identify the key approximations governing its performance. Our analysis proceeds by first establishing the numerical reliability of our computational approach, then dissecting the theoretical sources of error, and finally presenting a systematic, element-specific optimization that leads to a significant  improvement in accuracy.

\subsection*{Numerical Validation and Basis-Set Efficiency}

We first compare our results obtained with \( g_0^{-1} = 10 \) against those from Ref.~\cite{LeMaitre2023a}, where SCFT calculations employed a large Gaussian basis set. A critical point of comparison is the basis-set efficiency. While Ref.~\cite{LeMaitre2023a} utilized 153 Gaussian functions, our implementation employs a modest set of 70 B-splines  {of order 7}. Despite this more compact representation, the computed total energies are in excellent agreement. This demonstrates the superior efficiency of B-splines in representing electronic wavefunctions and densities, attributable to their flexibility and ability to capture rapid oscillations near the nucleus and asymptotic behavior at large radii with fewer degrees of freedom.

We rigorously tested the numerical convergence of our B-spline basis by increasing the number of splines beyond 70. As expected, no significant improvement in the total energy was observed. This confirms that the discrepancies between our SCFT results and the HF references are not due to numerical limitations of the basis set but are intrinsic to the theoretical approximations of the SCFT framework itself. Therefore, any pathway to improved accuracy must involve a refinement of the theory, not a mere expansion of the computational basis.

\subsection*{Decomposition of Theoretical Approximations}

To pinpoint the origin of the remaining errors and guide theoretical improvement, we examine the three core approximations in SCFT: (1) the \emph{shell approximation}, (2) the \emph{spherical averaging approximation}, and (3) the \emph{Pauli approximation}. The impact of the shell approximation has been analyzed in detail elsewhere~\cite{LeMaitre2023a}. Here, we focus on the latter two.

For hydrogen and helium, which are intrinsically spherically symmetric and contain no Pauli repulsion between like-spin electrons (H has one electron, He has a closed 1s shell), the SCFT formalism recovers the exact HF energy. This serves as a critical benchmark, confirming the soundness of the numerical implementation for simple cases. The elements lithium and beryllium are also spherically symmetric in their ground states, and the minor deviations observed (\(\sim\)0.5\% for Be with \( g_0^{-1}=10 \)) can be attributed to the Pauli approximation, as spherical symmetry is preserved.

For elements from boron to neon, which begin to develop non-spherical p-orbital densities, one might expect the spherical averaging approximation to introduce error. However, prior analysis by Chowdhury and Perdew~\cite{Chowdhury2021} indicates that the error introduced by spherically averaging the density for total energy calculations is remarkably small. Consequently, the primary source of the growing deviations observed with the constant prescription \( g_0^{-1} = 10 \), reaching \(\sim\)2.6\% for Ne, must be assigned to the limitations of the Pauli approximation. This approximation, which models the repulsion between same-spin electrons via a simplified potential, appears to lack the necessary scaling for increasingly complex, multi-electron systems.

This interpretation aligns with the observations in Ref.~\cite{LeMaitre2023a}, where a constant \( g_0^{-1} \) yielded reasonable accuracy up to krypton (deviations \(\sim\)3\%) but was noted to be improvable. The authors suggested that tuning \( g_0^{-1} \) as a function of \( Z \) could, in principle, reproduce HF results more closely. Our results strongly corroborate this: as shown in Table~\ref{tab:binding_energies}, the deviation for Xe with \( g_0^{-1}=10 \) exceeds 6\%, unequivocally demonstrating the breakdown of a constant parameter for heavy atoms.

\subsection*{Systematic Optimization and a   Z-Dependent Scaling}

Motivated by this analysis, we undertook a systematic, element-by-element optimization to determine the optimal \( g_0^{-1} \) value for each atom from H to Xe. For each element \( Z \), we minimized the difference between the SCFT free energy and the reference HF free energy~\cite{hf} by varying \( g_0^{-1} \). The resulting set of 54 optimal values was then subjected to a least-squares fitting procedure, which yielded the remarkably simple scaling relation:
\[
g_0^{-1} = \frac{192}{Z}.
\]

The binding energies calculated using this \( Z \)-dependent parameter are presented in the final columns of Table~\ref{tab:binding_energies}. The improvement is systematic. The previously observed monotonic increase in error with atomic number is eliminated. Instead, deviations are drastically reduced across the entire periodic table, consistently remaining below 0.7\% and often falling to 0.1\% or less. For many mid- to high-$Z$ elements, the agreement is within 0.01\%, effectively reproducing HF-level accuracy. This demonstrates that the primary failing of the simple Pauli approximation in SCFT is its lack of proper scaling with $Z$, which is elegantly corrected by the inverse proportionality \( g_0^{-1} \propto 1/Z \).

\subsection*{Limitations on Density Profiles and Outlook}

\begin{figure*}[h!]
\ffigbox{
\begin{tabular}{@{}c@{}c@{}}
\includegraphics[width=0.45\linewidth]{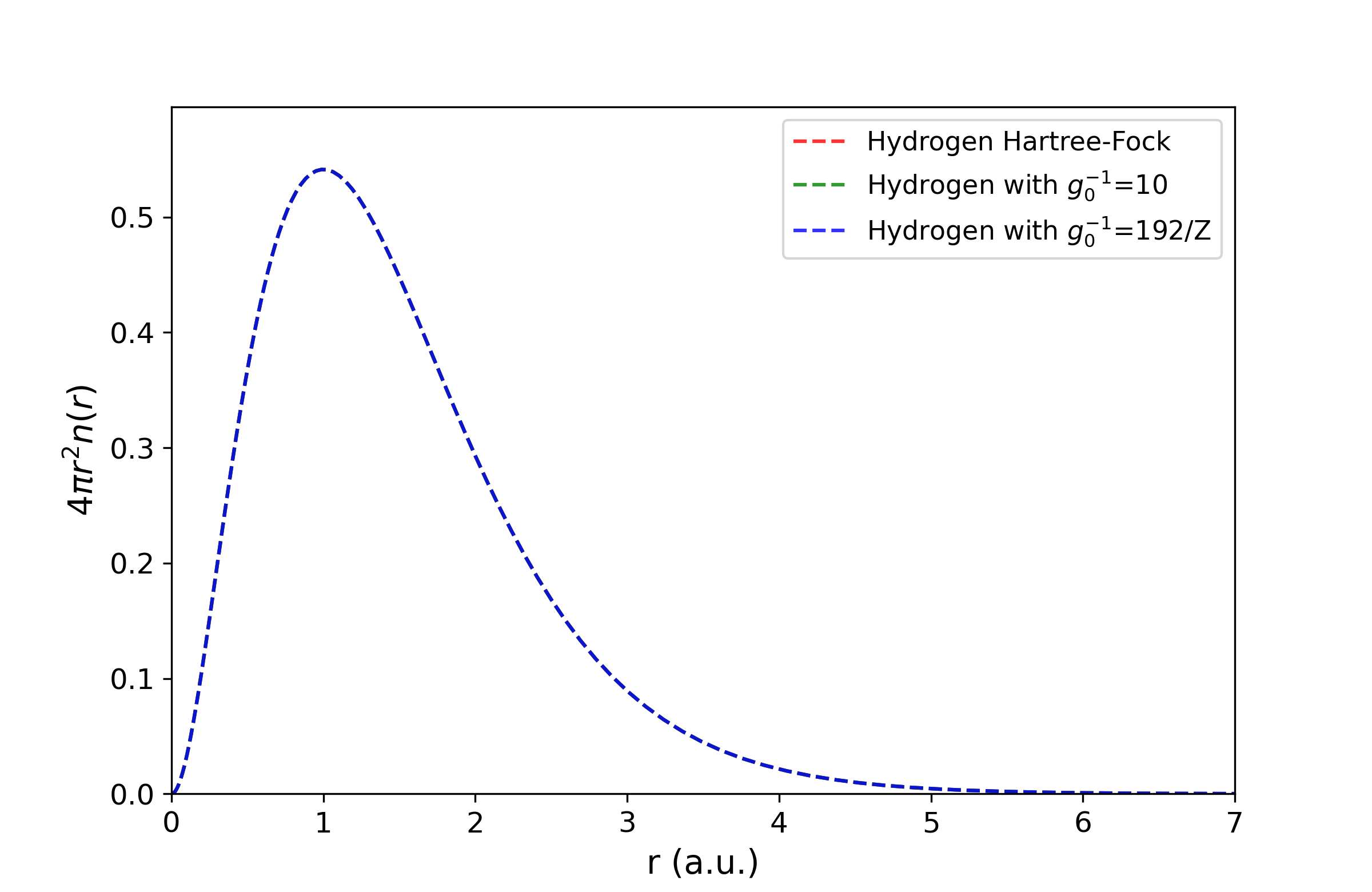} &
\includegraphics[width=0.45\linewidth]{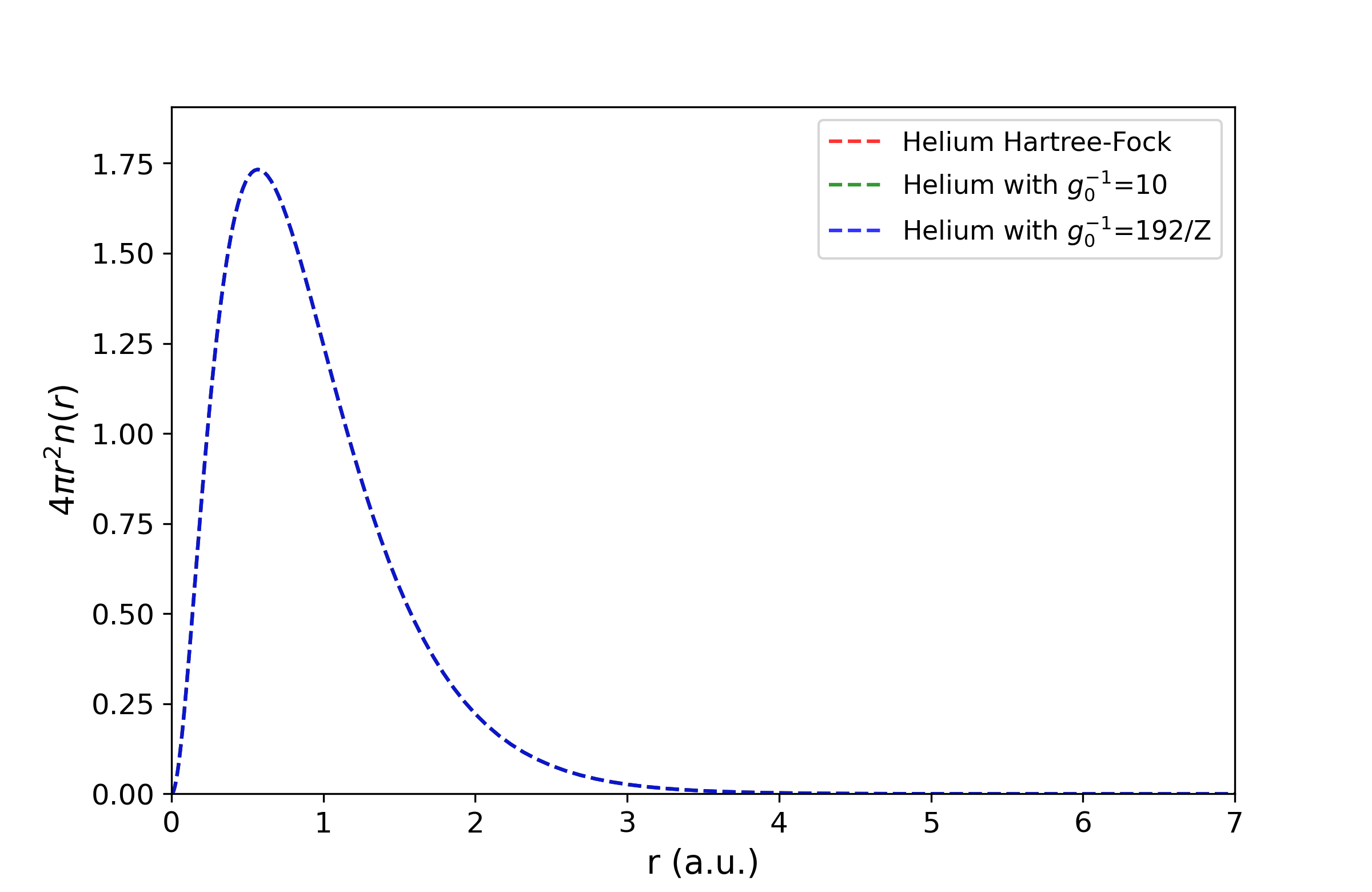} \\
(a) & (b) \\
\includegraphics[width=0.45\linewidth]{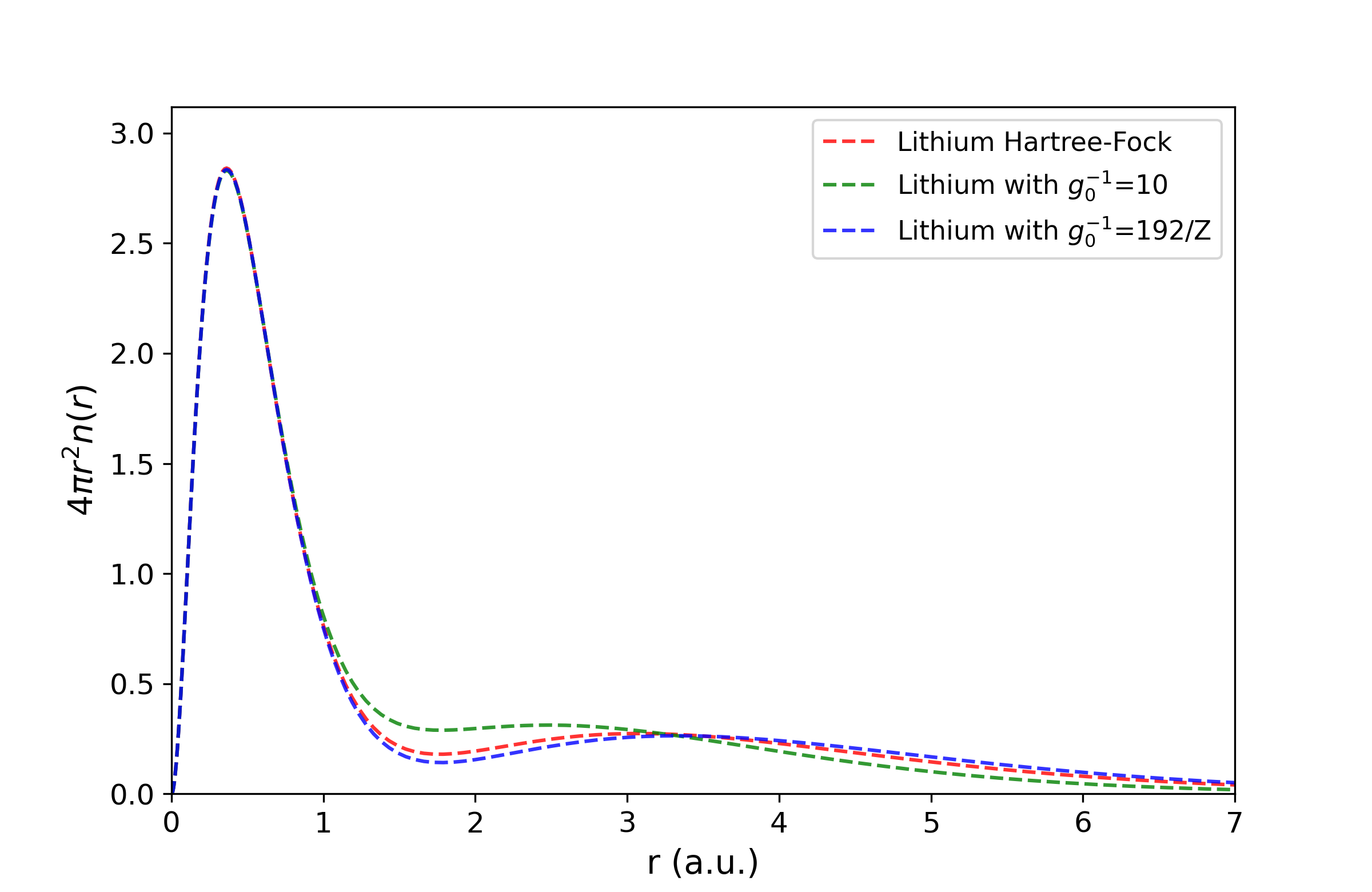} &
\includegraphics[width=0.45\linewidth]{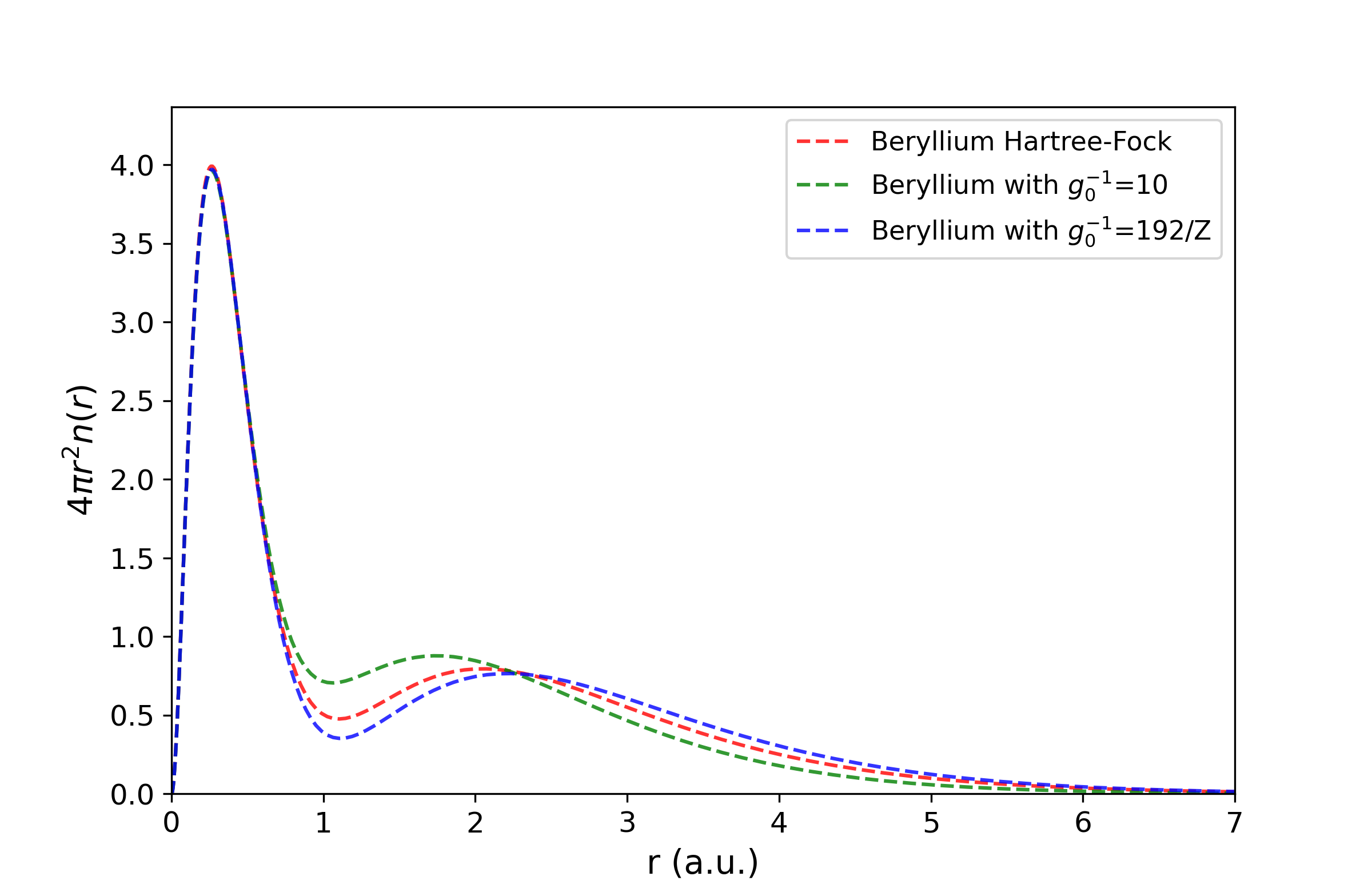} \\
(c) & (d)
\end{tabular}
}{
\caption{Plots of radial electron densities calculated using Hartree–Fock taken from Ref.~\cite{hf}, self-consistent field theory with $g_0^{-1}=10$ and self-consistent field theory with $g_0^{-1}=192/Z$ for (a) H, (b) He, (c) Li, (d) Be.}
\label{aaa}}
\end{figure*}

\begin{figure*}[h!]
\ffigbox{
\begin{tabular}{@{}c@{}c@{}}
\includegraphics[width=0.48\linewidth]{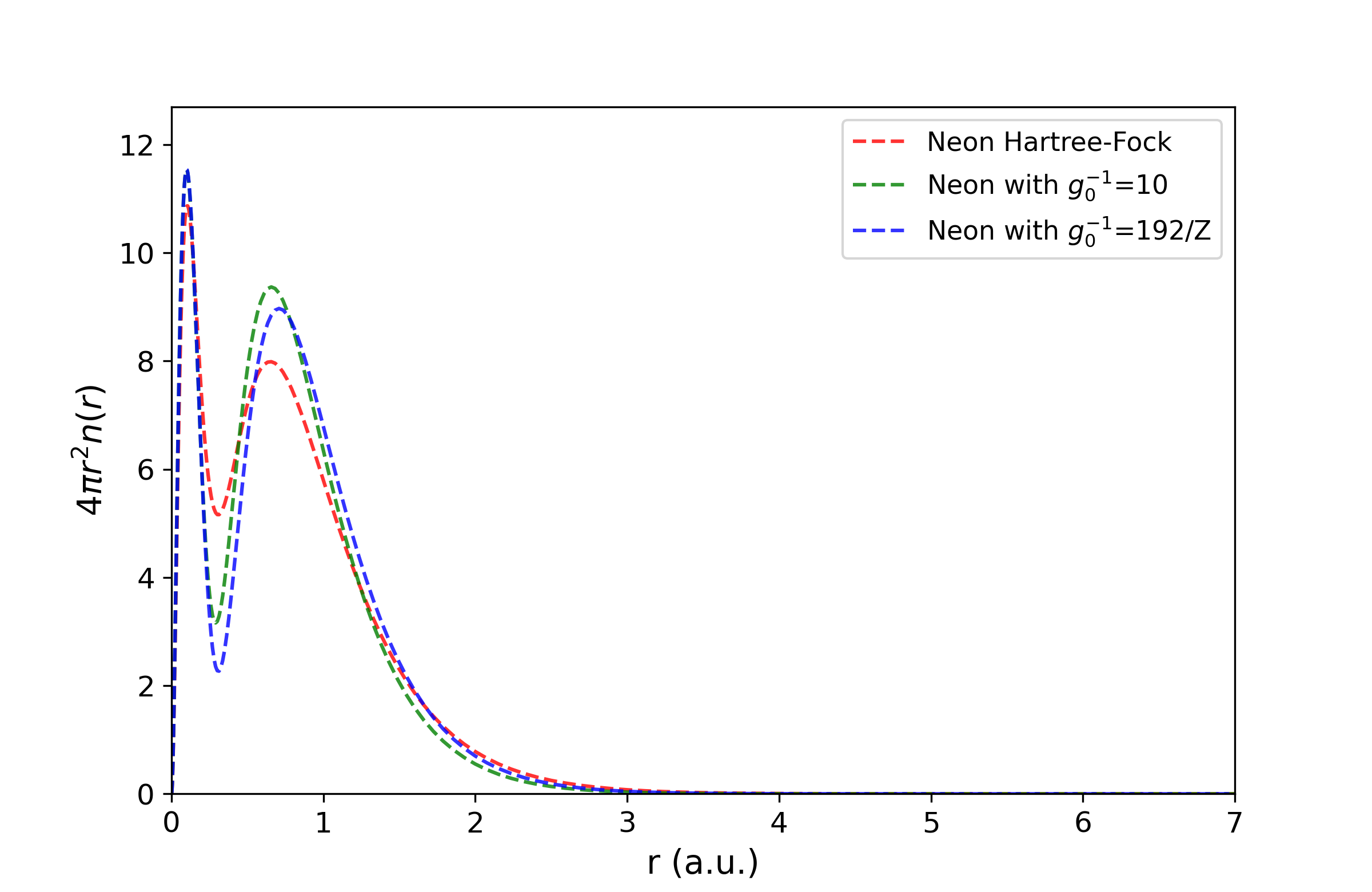} &
\includegraphics[width=0.48\linewidth]{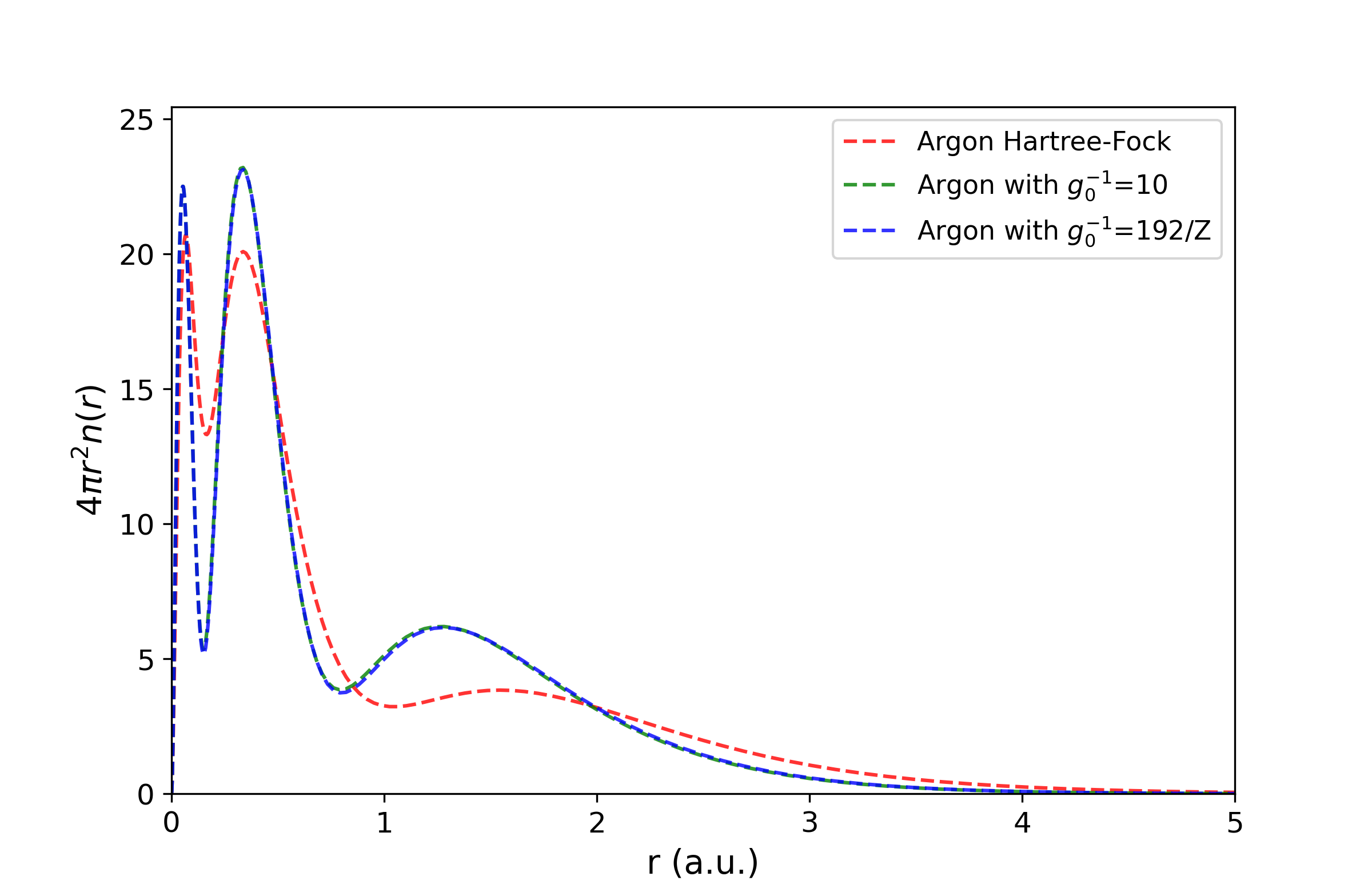} \\
(a) Ne & (b) Ar \\[6pt]
\includegraphics[width=0.48\linewidth]{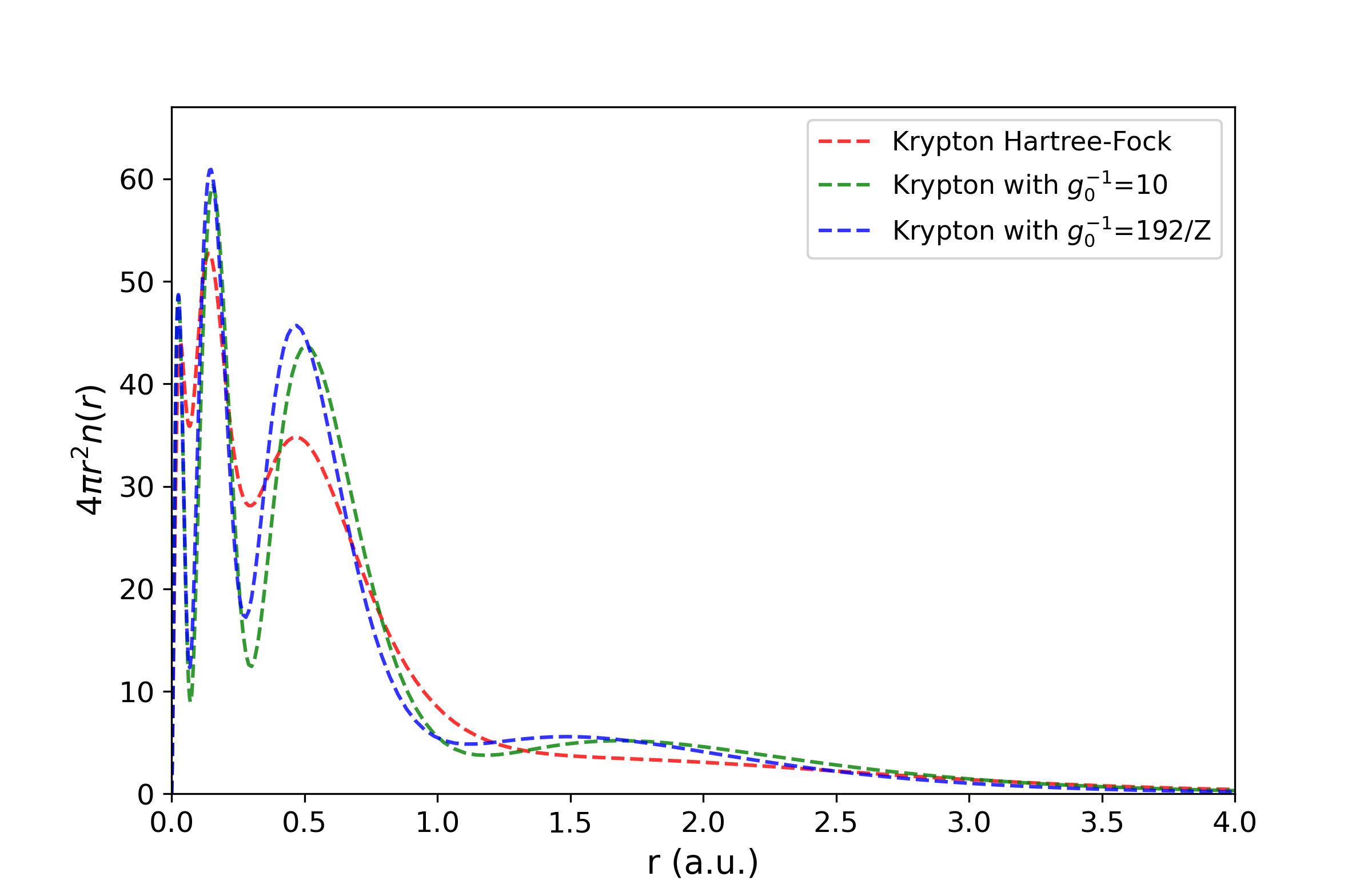} &
\includegraphics[width=0.48\linewidth]{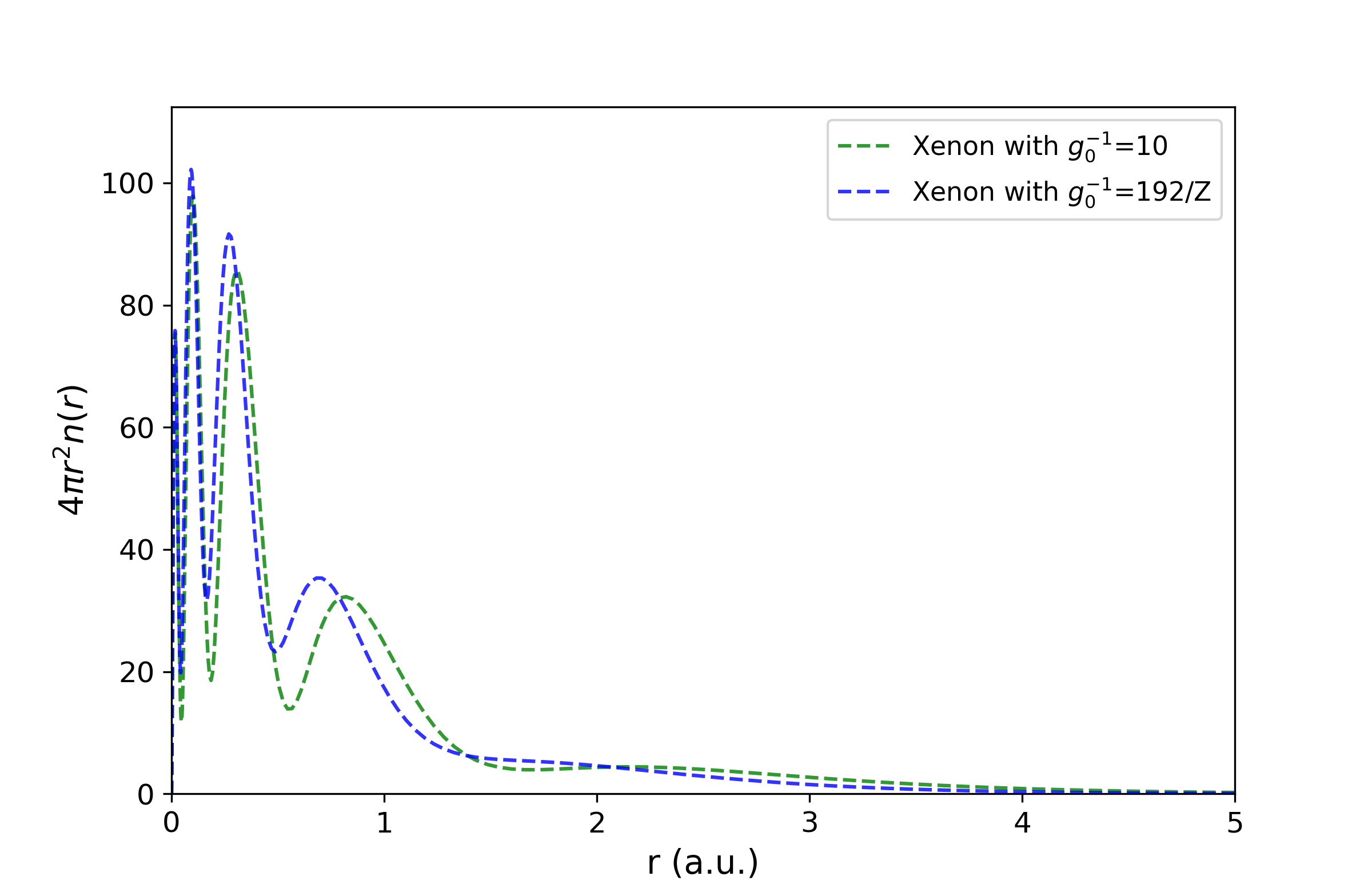} \\
(c) Kr & (d) Xe
\end{tabular}
}{
\caption{Plots of radial electron densities calculated using Hartree–Fock taken from Ref.~\cite{hf}, self-consistent field theory with $g_0^{-1}=10$ and self-consistent field theory with $g_0^{-1}=192/Z$ for (a) Ne, (b) Ar, (c) Kr, (d) Xe. Note that HF values for Xe were not available in~\cite{hf}.}
\label{bbb}}
\end{figure*}

While the \( Z \)-dependent scaling achieves excellent accuracy for total energies, it is important to note its scope. This empirical scaling corrects a global, integrated property (the total energy) but does not fully rectify the \emph{local} structure of the approximate Pauli potential. As a result, while electron density profiles for light atoms (H through Be) show excellent agreement with HF, noticeable discrepancies persist for the radial densities of heavier atoms like Ne and Kr, as illustrated in Figs.~\ref{aaa} and \ref{bbb}. This is expected, as the functional form of the Pauli approximation differs from the exact exchange potential by more than just a simple scalar multiplier. Improving the \emph{shape} of the Pauli potential, rather than just its overall strength, represents the next frontier for advancing the SCFT formalism towards a fully orbital-free density functional theory that accurately reproduces both energies and densities.

\section{Conclusion}

In this work, we have significantly advanced the application of the self-consistent field theory to the quantum many-body problem by systematically addressing two fundamental challenges: the approximate treatment of Pauli repulsion and the choice of an optimal numerical basis.

First, we have demonstrated that the accuracy of SCFT is critically dependent on the scaling of its Pauli repulsion parameter, \( g_0^{-1} \). The conventional approach of employing a constant value, while reasonably accurate for light atoms, fails for heavier elements, leading to systematic errors exceeding 6\%. Through a systematic, element-by-element optimization, we derived a simple \( Z \)-dependent scaling, \( g_0^{-1} = 192/Z \). This modification improves agreement with Hartree-Fock reference energies from H to Xe, reducing deviations to well below 1\%.

Second, we have established B-spline basis functions as an optimal numerical framework for solving the SCFT equations. Compared to the Gaussian basis sets traditionally used, our B-spline implementation achieved comparable accuracy with fewer than half the basis functions. The inherent properties of B-splines, local support, smooth polynomial nature, and tunable resolution-provide an optimal balance between numerical efficiency and accuracy.

Taken together, these innovations allow SCFT to produce predictions in near-perfect agreement with Hartree-Fock energies while preserving the framework's inherent advantages: computational simplicity, orbital-free formalism, and favorable scaling. This progress validates SCFT as a robust and practical first-principles method for atomic systems and clarifies the path for its future development. The remaining discrepancies in electron density profiles for heavier elements point directly to the next necessary refinement: moving beyond a simple scalar Pauli potential to one with a more accurate non-local or density-gradient-dependent form.

% These developments establish a solid foundation for extending SCFT to more complex quantum systems where orbital-free methods offer a compelling computational advantage. The demonstrated scalability and accuracy pave the way for applications to molecular systems, where electron correlation and chemical bonding present new challenges, and to condensed matter environments, where the efficient treatment of extended systems is paramount.

\section*{Acknowledgement}
This project was carried out during the authors' stay at the Perimeter Institute for Theoretical Physics, Waterloo. VB gratefully acknowledges financial support from the University of Waterloo and the Perimeter Institute for the opportunity to work as an IGVS student at the University of Waterloo. BA acknowledges support from the Perimeter Institute.

%\bibliography{refs} 


%\input{Tables/Table III}
%\input{Figures/5. Selected1}
%\input{Figures/5. Selected2}

\begin{longtable*}[!h]{p{0.5cm}p{1cm}p{2.5cm}p{3cm}p{2.0cm}p{2cm}p{2.5cm}p{1cm}}
    \caption{Atomic binding energies (in atomic units) for elements from hydrogen to xenon, computed using SCFT and B-splines {(order 7)} with two choices of \( g_0^{-1} \): a constant value of 10 and a $Z$-dependent value \( g_0^{-1} = \frac{192}{Z} \). Hartree-Fock (HF) reference values and deviations of SCFT results from HF are included. Values in brackets are from Ref.~\cite{LeMaitre2023a}, obtained using Gaussian basis functions.} \label{tab:binding_energies} \\
    
    \hline
    Z & Atom & Electronic configuration & Hartree-Fock\cite{hf}  & SCFT free energy ($g_0^{-1} = 10$) & \% difference & SCFT free energy ($g_0^{-1} = 192/Z$) & \% difference \\ 
    \hline
    \endfirsthead
    
    % Header for subsequent pages
    \multicolumn{8}{c}{{\tablename\ \thetable{} -- continued from previous page}} \\
    \hline
    Z & Atom & Electronic configuration & Hartree-Fock\cite{hf}  & SCFT free energy ($g_0^{-1} = 10$) & \% difference & SCFT free energy ($g_0^{-1} = 192/Z$) & \% difference \\ 
    \hline
    \endhead
    
    % Footer for page breaks
    \hline
    \multicolumn{8}{r}{{Continued on next page}} \\
    \endfoot
    
    % Footer for last page
    \hline
    \endlastfoot

        1 & H & 1s$^1$(1) & 0.5 & .5000000000 (0.49999998) & 1.2E-9 & 0.5 & 1.2E-9 \\ 
        2 & He & 1s(2) & 2.861679993 & 2.861679996 (2.86168) & 1.2E-7 & 2.861679996 & 1.2E-7 \\ 
        3 & Li & [He]2s(1) & 7.432726924 & 7.468422201 (7.46842) & 0.48 & 7.420324425 & 0.17 \\ 
        4 & Be & [He]2s(2) & 14.57302313 & 14.70219466 (14.70219) & 0.89 & 14.50238653 & 0.48 \\ 
        5 & B & [He]2s(3) & 24.52906069 & 24.90399685 (24.904) & 1.53 & 24.44312625 & 0.35 \\ 
        6 & C & [He]2s(4) & 37.6886189 & 38.40322533 (38.40323) & 1.90 & 37.58383716 & 0.28 \\ 
        7 & N & [He]2s(5) & 54.40093415 & 55.52626881 (55.52627) & 2.07 & 54.27219573 & 0.24 \\ 
        8 & O & [He]2s(6) & 74.8093984 & 76.59988851 (76.59989) & 2.39 & 74.86201223 & 0.07 \\ 
        9 & F & [He]2s(7) & 99.40934928 & 101.9529490 (101.95295) & 2.56 & 99.71292817 & 0.31 \\ 
        10 & Ne & [He]2s(8) & 128.547098 & 131.9173464 (131.91735) & 2.62 & 129.1901237 & 0.50 \\ 
        11 & Na & [Ne]3s(1) & 161.8589113 & 165.5248018 (165.5246) & 2.26 & 162.4297251 & 0.35 \\ 
        12 & Mg & [Ne]3s(2) & 199.6146361 & 203.3441068 (203.3441) & 1.87 & 199.9763468 & 0.18 \\ 
        13 & Al & [Ne]3s(3) & 241.876707 & 245.5075707 (245.5076) & 1.50 & 241.9866211 & 0.05 \\ 
        14 & Si & [Ne]3s(4) & 288.8543622 & 292.1387109 (292.1387) & 1.14 & 288.612136 & 0.08 \\ 
        15 & P & [Ne]3s(5) & 340.7187806 & 343.3592771 (343.3593) & 0.77 & 340.0043158 & 0.21 \\ 
        16 & S & [Ne]3s(6) & 397.5048955 & 399.2910599 (399.2911) & 0.45 & 396.3155226 & 0.30 \\ 
        17 & Cl & [Ne]3s(7) & 459.4820719 & 460.0565780 (460.0566) & 0.13 & 457.6994043 & 0.39 \\ 
        18 & Ar & [Ne]3s(8) & 526.8175122 & 525.7794077 (525.7794) & 0.20 & 524.3110578 & 0.48 \\ 
        19 & K & [Ar]4s(1) & 599.1647831 & 595.9667015 (595.9661) & 0.53 & 595.6902789 & 0.58 \\ 
        20 & Ca & [Ar]4s(2) & 676.7581817 & 670.9223571 (670.9221) & 0.86 & 672.1626061 & 0.68 \\ 
        21 & Sc & [Ar]3s(1)4s(2) & 759.7357123 & 752.0402271 (752.04) & 1.01 & 755.1590305 & 0.60 \\ 
        22 & Ti & [Ar]3s(2)4s(2) & 848.4059907 & 838.6136330 (838.6134) & 1.15 & 844.0063078 & 0.52 \\ 
        23 & V & [Ar]3s(3)4s(2) & 942.8843308 & 930.7715505 (930.7714) & 1.28 & 938.8659616 & 0.43 \\ 
        24 & Cr & [Ar]3s(5)4s(1) & 1043.356368 & 1030.160861 (1030.1603) & 1.26 & 1041.458451 & 0.18 \\ 
        25 & Mn & [Ar]3s(5)4s(2) & 1149.866243 & 1132.364797 (1132.365) & 1.52 & 1147.276549 & 0.23 \\ 
        26 & Fe & [Ar]3s(6)4s(2) & 1262.443656 & 1242.066018 (1242.066) & 1.61 & 1261.158429 & 0.10 \\ 
        27 & Co & [Ar]3s(7)4s(2) & 1381.414542 & 1357.883272 (1357.883) & 1.70 & 1381.71421 & 0.02 \\ 
        28 & Ni & [Ar]3s(8)4s(2) & 1506.870896 & 1479.952401 (1479.953) & 1.79 & 1509.112696 & 0.15 \\ 
        29 & Cu & [Ar]3s(10)4s(1) & 1638.963723 & 1610.738290 (1610.739) & 1.72 & 1645.972851 & 0.43 \\ 
        30 & Zn & [Ar]3s(10)4s(2) & 1777.848102 & 1743.396015 (1743.398) & 1.94 & 1785.120146 & 0.41 \\ 
        31 & Ga & [Ar]3s(10)4s(3) & 1923.261001 & 1881.817842 (1881.819) & 2.15 & 1930.664496 & 0.38 \\ 
        32 & Ge & [Ar]3s(10)4s(4) & 2075.359726 & 2026.072322 (2026.074) & 2.37 & 2082.701738 & 0.35 \\ 
        33 & As & [Ar]3s(10)4s(5) & 2234.238647 & 2176.222766 (2176.225) & 2.60 & 2241.323554 & 0.32 \\ 
        34 & Se & [Ar]3s(10)4s(6) & 2399.867604 & 2332.331282 (2332.335) & 2.81 & 2406.619927 & 0.28 \\ 
        35 & Br & [Ar]3s(10)4s(7) & 2572.441325 & 2494.461277 (2494.468) & 3.03 & 2578.679651 & 0.24 \\ 
        36 & Kr & [Ar]3s(10)4s(8) & 2752.054969 & 2662.677326 (2662.684) & 3.25 & 2757.590792 & 0.20 \\ 
        37 & Rb & [Kr]5s(1) & 2938.357442 & 2836.610874 & 3.46 & 2942.985556 & 0.16 \\ 
        38 & Sr & [Kr]5s(2) & 3131.545674 & 3016.48958 & 3.67 & 3135.169185 & 0.12 \\ 
        39 & Y & [Kr]4s(1)5s(2) & 3331.684158 & 3203.188183 & 3.86 & 3334.977735 & 0.10 \\ 
        40 & Zr & [Kr]4s(2)5s(2) & 3538.995053 & 3396.217633 & 4.03 & 3541.998336 & 0.08 \\ 
        41 & Nb & [Kr]4s(4)5s(1) & 3753.597716 & 3596.535207 & 4.18 & 3757.2735 & 0.10 \\ 
        42 & Mo & [Kr]4s(5)5s(1) & 3975.549487 & 3802.510643 & 4.35 & 3979.090146 & 0.09 \\ 
        43 & Tc & [Kr]4s(5)5s(2) & 4204.788722 & 4013.956733 & 4.54 & 4207.212145 & 0.06 \\ 
        44 & Ru & [Kr]4s(7)5s(1) & 4441.539471 & 4234.137058 & 4.67 & 4445.244296 & 0.08 \\ 
        45 & Rh & [Kr]4s(8)5s(1) & 4685.881686 & 4459.915771 & 4.82 & 4689.765136 & 0.08 \\ 
        46 & Pd & [Kr]4s(10) & 4937.921004 & 4693.454096 & 4.95 & 4943.234394 & 0.11 \\ 
        47 & Ag & [Kr]4s(10)5s(1) & 5197.698452 & 4931.698414 & 5.12 & 5202.162368 & 0.09 \\ 
        48 & Cd & [Kr]4s(10)5s(2) & 5465.133119 & 5176.295637 & 5.29 & 5468.485309 & 0.06 \\ 
        49 & In & [Kr]4s(10)5s(3) & 5740.169136 & 5427.304592 & 5.45 & 5742.286951 & 0.04 \\ 
        50 & Sn & [Kr]4s(10)5s(4) & 6022.931678 & 5684.77991 & 5.61 & 6023.635042 & 0.01 \\ 
        51 & Sb & [Kr] 4s(10)5s(5) & 6313.485304 & 5948.774597 & 5.78 & 6312.592021 & 0.01 \\ 
        52 & Te & [Kr]4s(10)5s(6) & 6611.784043 & 6219.335507 & 5.94 & 6609.21898 & 0.04 \\ 
        53 & In & [Kr]4s(10)5s(7) & 6917.980881 & 6496.498418 & 6.09 & 6913.577479 & 0.06 \\ 
        54 & Xe & [Kr]4s(10)5s(8) & 7232.138349 & 6780.286886 & 6.25 & 7225.729986 & 0.09 \\ \hline

\end{longtable*}

\end{document}